\documentclass[aps,prl,reprint]{revtex4-1}
\usepackage{graphicx}
\usepackage{dcolumn}
\usepackage{bm}
\usepackage{lipsum}
\usepackage{subcaption}
\usepackage{ragged2e}
\usepackage{balance}
\DeclareCaptionJustification{justified}{\justifying}
\usepackage{float}
\usepackage{amsmath, amsthm, amssymb, amsfonts}
\usepackage{color}

\begin{document}


\title{Jamming of Deformable Polygons}


\author{Arman Boromand$^{1,2}$, Alexandra Signoriello$^{2,3}$, Fangfu Ye$^{1,4}$, Corey S. O'Hern$^{2,3,5,6}$, and Mark D. Shattuck$^{7}$}

\affiliation{$^{1}$Beijing National Laboratory for Condensed Matter Physics and CAS Key Laboratory of Soft Matter Physics, Institute of Physics, Chinese Academy of Sciences, Beijing, China}
\affiliation{$^{2}$Department of Mechanical Engineering and Materials Science, Yale University, New Haven, Connecticut, 06520, USA} 
\affiliation{$^{3}$Program in Computational Biology and Bioinformatics, Yale University, New Haven, Connecticut, 06520, USA}
\affiliation{$^{4}$School of Physical Sciences, University of Chinese Academy of Sciences, Beijing, China}
\affiliation{$^{5}$Department of Physics, Yale University, New Haven, Connecticut, 06520, USA}
\affiliation{$^{6}$Department of Applied Physics, Yale University, New Haven, Connecticut, 06520, USA}
\affiliation{$^{7}$Benjamin Levich Institute and Physics Department,
The City College of the City University of New York, New York, New York 10031, USA\\}


\date{\today}

\begin{abstract}
There are two main classes of physics-based models for two-dimensional 
cellular materials: packings of repulsive disks and the vertex model.  These
models have several disadvantages. For example, disk interactions are
typically a function of particle overlap, yet the model assumes that
disks remain circular during overlap. The shapes of the cells can vary
in the vertex model, however, the packing fraction is fixed at
$\phi=1$. Here, we describe the deformable particle model (DPM), where each
particle is a polygon composed of a large number of vertices.  The
total energy includes three terms: two quadratic terms to penalize
deviations from the preferred particle area $a_0$ and perimeter $p_0$
and a repulsive interaction between DPM polygons that penalizes
overlaps. We performed simulations to study jammed DPM
packings as a function of asphericity, ${\cal A} =
p_0^2/4\pi a_0$. We show that the packing fraction at jamming onset
$\phi_J({\cal A})$ grows with increasing ${\cal A}$, reaching
confluence at ${\cal A}^* \approx 1.16$.  ${\cal A}^*$ corresponds to
the value at which DPM polygons completely fill the cells obtained
from surface-Voronoi tessellation. Further, we find that
DPM polygons develop invaginations for ${\cal A} > {\cal A}^*$ with
excess perimeter that grows linearly with ${\cal A}-{\cal A}^*$. We also
confirm that DPM packings are solid-like for ${\cal A} > {\cal A}^*$ 
and ${\cal A} < {\cal A}^*$. 
\end{abstract}

\maketitle

There are many physical systems that can be modeled as packings of
discrete, deformable particles, including cell monolyers, developing
embryos, foams, and emulsions~\cite{Martind_development,sadatidiff,kasza, ang_glass_pnas,vanhecke_epl,bruj_phys_A}. A spectrum of models with varying
degrees of complexity have been employed to study these systems.
Perhaps the simplest model involves packings of disk-shaped particles
that interact via repulsive forces~\cite{ohern, ohern2, szabo,henkes}.  The power of this model is
its simplicity and the ability to study a range of packing
fractions $\phi$ from below jamming, where particles are not in contact, to
jamming onset, where nearly all particles are at contact, to above
jamming, where the particles are over-compressed. However, in this
model, forces between particles are generated via overlaps,
and the particles remain spherical during overlap, which is
unphysical. In contrast, the vertex model~\cite{farhadifar, staple} in two spatial dimensions
(2D) employs deformable polygons (with a relatively small number of
vertices, but different polygonal shapes), with no particle overlaps,
to study the structural and mechanical properties of cell monolayers.
However, the vertex model only considers confluent systems with
$\phi=1$, and thus it cannot describe inter-cellular
space.\par

Disk-packing models allow us to study the onset of jamming of 2D cellular
materials as a function of packing fraction, whereas the vertex model
allows us to study the onset of jamming as a function of particle
shape, e.g. the asphericity, ${\cal A} = p^2/4 \pi a$, where $p$ and
$a$ are the perimeter and area of the particles~\cite{binature,biprx}.  Here, we
introduce the deformable particle model (DPM), which enables us to
vary {\it both} the packing fraction and particle shape.
In 2D, the DPM is a polygon with a large number of vertices, which
enables modeling of particle deformation.  The total energy of a
collection of DPM polygons includes three terms. Two quadratic terms
for each polygon to penalize deviations from the preferred area and
perimeter and a repulsive contact interaction between pairs of
deformable polygons to penalize overlaps.

We performed simulations to study jamming onset in
DPM packings and found several key results. First, we show that the
packing fraction at jamming onset $\phi_J({\cal A})$ increases with
${\cal A}$, starting at $\phi_J \approx 0.81$ or $\approx 0.88$ for
monodisperse disks (${\cal A}=1$), depending on the roughness of the
particles, and reaching $\phi_J=1$ for ${\cal A} \geq {\cal A}^*$,
where ${\cal A}^* \approx 1.16$.  We find similar results for
$\phi_J({\cal A})$ in packings of bidisperse deformable
polygons.  We show that
${\cal A}^*$ corresponds to the value at which DPM polygons
completely fill the cells obtained from Voronoi tessellation. Further,
for ${\cal A} > {\cal A}^*$, the deformable polygons develop
invaginations, which grow with ${\cal A}-{\cal A}^*$. We show that the
distributions of Voronoi areas for jammed DPM packings follow
$k$-gamma distributions for all ${\cal A}$, which is a hallmark of
jamming in systems composed of rigid
particles~\cite{aste_pre_emergence}. By calculating the static shear
modulus, we also confirm that DPM packings are solid-like for all ${\cal A}$.

For the DPM, each ``particle" is a collection of $N_v$ vertices that
form an $N_v$-sided deformable polygon.  (See Fig.~\ref{fig.1}.) Each
polygon has $N_v$ edges indexed by $i=1,\ldots,N_v$. To
ensure that each particle remains a polygon, adjacent vertices are
connected via linear springs, with spring constant $k_l$ and
equilibrium length $l_0 = p_0/N_v$, where $p_0$ is the preferred perimeter of
the polygon.  For reference, ${\cal A}$ for a rigid (regular)
polygon with $N_v$ vertices is ${\cal A}_v = N_{v}\tan(\pi / N_{v})/
\pi$, which reduces to ${\cal A}_v=1$ when $N_{v}\to\infty$.

The total energy, $U$, for the DPM also includes a quadratic term
that penalizes deviations of the polygon area $a$
from the reference value $a_0$, which models particle elasticity.
In addition, we include a pairwise, repulsive interaction
energy, $U_{\rm int}$, to prevent overlaps between polygons. The total
energy for $N$ deformable polygons is therefore
\begin{eqnarray}
\label{tote}
U & = & \sum_{m=1}^N \sum_{i=1}^{N_v} \frac{k_l}{2} (l_{mi}-l_0)^2 +
\sum_{m=1}^{N} \frac{k_a}{2}(a_m-a_0)^2\\
& + & U_{\rm int}, \nonumber
\nonumber
\end{eqnarray}
where $l_{mi}$ is the length of the $i$th edge of polygon $m$ and
$k_a$ is the spring constant for the quadratic term in area, which is
proportional to the polygon's compressibility.

We implement two methods for calculating the repulsive interactions
between deformable polygons.  For the {\it rough surface method}, we
fix disks with diameter $\delta=l_0=1$ at each 
polygon vertex (Fig.~\ref{fig.1} (a) and (b)). In this case, the repulsive interactions
are obtained by summing up repulsive linear 
spring interactions between overlapping disks on contacting polygons:     
\begin{eqnarray}
\label{int}
U_{\rm int} & = & \sum_{m=1}^N \sum_{n>m}^N\sum_{j=1}^{N_v}\sum_{k=1}^{N_v}\frac{k_r}{2}(\delta-|
\textbf{v}_{mj}-\textbf{v}_{nk}|)^2\\
& \times &\Theta(\delta-|\textbf{v}_{mj}-\textbf{v}_{nk}|),
\nonumber
\end{eqnarray}
where $k_r$ gives the strength of the repulsive interactions, ${\bf
  v}_{mj}$ is the position of the $j$th vertex in polygon $m$ and
$\Theta(.)$ is the Heaviside step function.  We also implemented a
{\it smooth surface method} by modeling the polygon edges as
circulo-lines (i.e.~the collection of points that are a fixed distance
from a line) with width $\delta$~\cite{kylejamming}. (See
Fig.~\ref{fig.1} (c) and (d).)  In this method, we again use
Eq.~\ref{int} for the repulsive interactions between polygons, except
the overlap ($\delta-|\textbf{v}_{mj}-\textbf{v}_{nk}|$) is replaced
by $\delta - d_{\rm min}$, where $d_{\rm min}$ is minimum distance
between the line segments $l_{mj}$ and $l_{nk}$ on contacting polygons
$m$ and $n$.  We set the ratios $k_l l_0^2/k_a = 10$ and $k_l/k_r = 1$;
other values of these parameters yield similar results near jamming
onset. Energies are measured in units of $k_a l_0^2$ below.

To generate static packings, we place polygons with random locations
and orientations in a square box with periodic boundary conditions and
$\phi = 0.2$.  We successively compress the system isotropically using
small packing fraction increments $d\phi<10^{-4}$ and minimizing $U$
after each compression step using over-damped molecular dynamics
simulations until the kinetic energy per particle $K/N < 10^{-20}$. We
use bisection with compression and decompression to identify jamming
onset, where the total energy per particle satisfies $0 < U/N <
10^{-16}$.

\begin{figure}[H]
\centering
\includegraphics[width=0.48\textwidth]{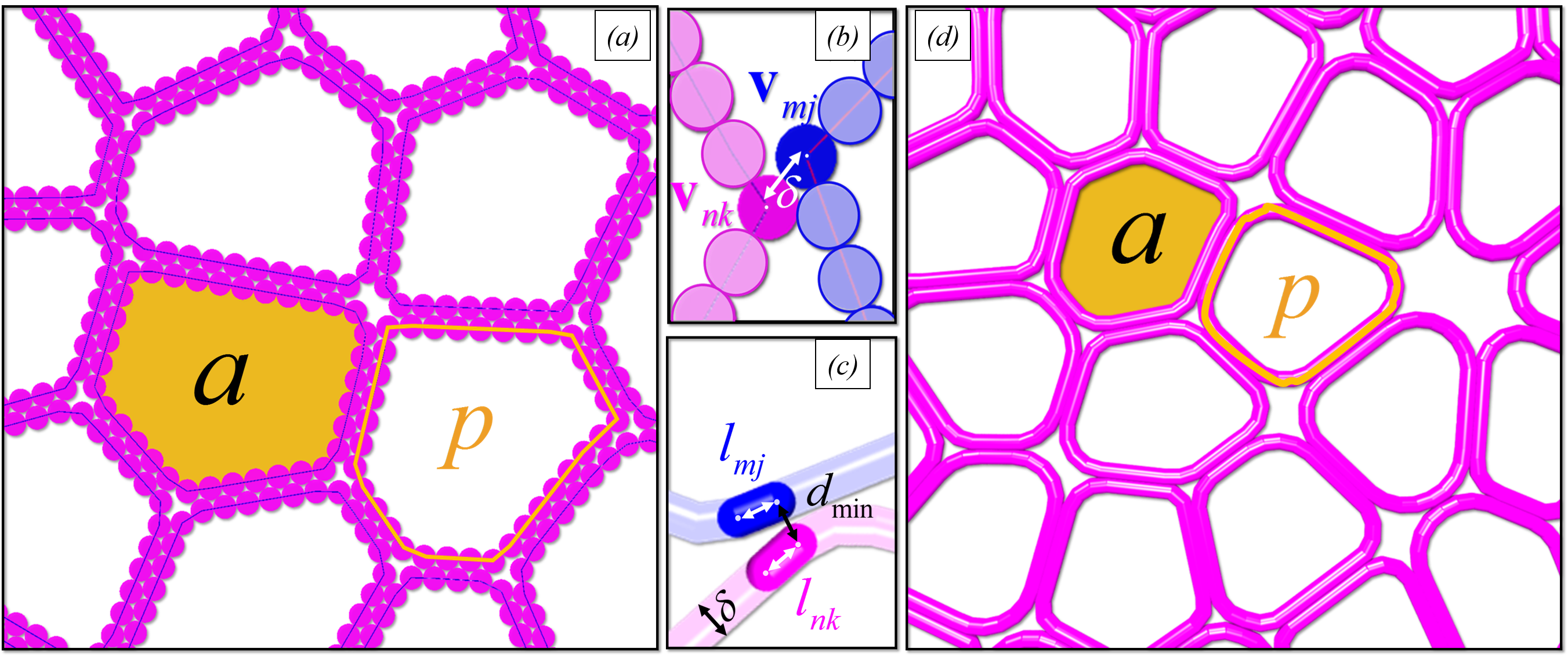}\label{fig.1}

\caption{Schematic of deformable polygons with
$N_{v}=34$ vertices (with the position of the $j$th vertex in the 
$m$th polygon given by ${\bf v}_{mj}$), area $a$, and perimeter $p$. 
${l}_{mj}=p/N_v$ is the line segment between vertices $j$ and $j+1$ in 
polygon $m$. We implemented 
two methods for modeling edges of deformable polygons. 
In (a) and 
(b), we show the 
rough surface method, where we fix the centers of disks with 
diameter $\delta$ at polygon vertices.
In (c) and (d), we 
show the smooth surface method, where we model polygon edges as 
circulo-lines with width $\delta$. $d_{\rm min}$ is minimum distance 
between line segments $l_{mj}$ and $l_{nk}$. 
}
\label{fig.1}
\end{figure}

We show the packing fraction at jamming onset $\phi_J$ (normalized by
the maximum packing fraction for each surface roughness model,
$\phi_{max}$) versus asphericity ${\cal A}/{\cal A}_v$ for $N=64$
deformable polygons in Fig.~\ref{fig.2} (a). Note that $\phi_{\rm max}
\approx 0.99$ and $0.95$ for the smooth and rough surface methods,
respectively, for $N_v=12$ and the maximum packing fraction for both
methods converges to $\phi_{\rm max} = 1$ as $N_v \rightarrow \infty$
[Fig.~\ref{fig.2} (b)].  $\phi_J/\phi_{\rm max} \approx 0.81$ ($0.88$)
for the rough (smooth) surface method near ${\cal A}/{\cal A}_v = 1$
and $\phi_J$ grows with increasing ${\cal A}/{\cal
  A}_v$. As expected, $\phi_J$ for the rough surface method in the
rigid-disk limit is smaller than that for the smooth surface method.
The results obtained near ${\cal A}=1$ are similar to previous results
for jammed packings of monodisperse, frictionless ($\phi_J \approx
0.88$-$0.89$~\cite{donev}) and frictional disks ($\phi_J \approx
0.8$~\cite{kondic}).  For ${\cal A}/{\cal A}_v > 1.02$,
$\phi_J/\phi_{\rm max}$ possess similar dependence on ${\cal A}$ for
the two surface roughness methods.  We also find similar results for
jammed packings of bidisperse DPM polygons (half large with $N_v=17$ and
half small with $N_v=12$ and perimeter ratio $r=1.4$).  As shown in
Fig.~\ref{fig.2} (b), the jammed packings become confluent with
$\phi_J \approx 1$ for ${\cal A} > {\cal A}^* \approx 1.16$ in the
large $N_v$ limit.

\begin{figure}[H]
\captionsetup[subfloat]{labelformat=empty,justification=raggedright ,singlelinecheck=false}
\includegraphics[trim={0.1cm 0cm 0.1cm 0.0cm},clip, width=0.5\textwidth]{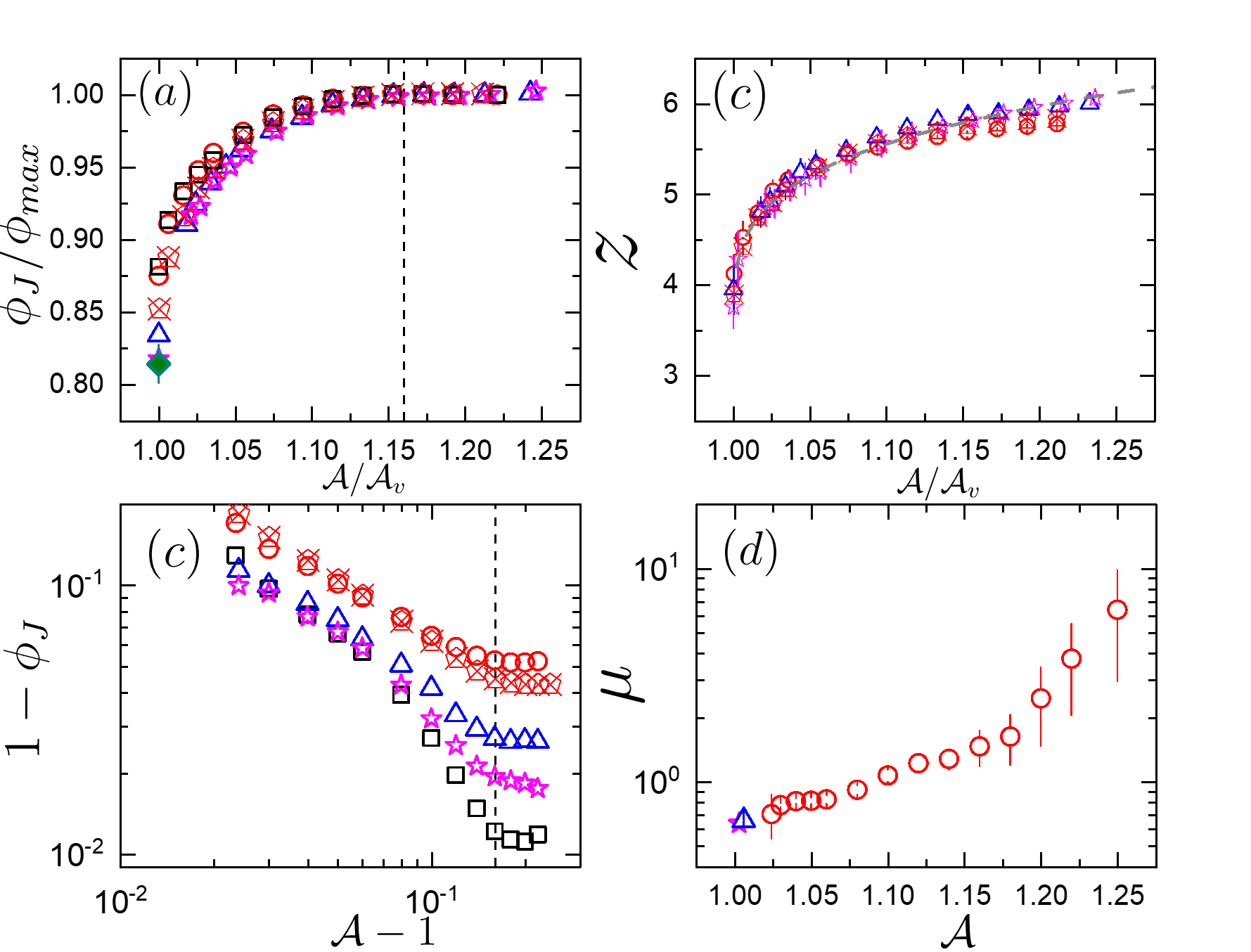}\label{fig.2}
\caption{(a) Packing fraction at jamming onset $\phi_J$
(normalized by the maximum packing fraction, $\phi_{\rm max}$, for each
surface roughness model), (b) the deviation of $\phi_J$ from the
confluent value, $1-\phi_J$, (c) coordination number $z$,  and (d)
average friction coefficient $\mu$ (for the rough surface model) for
DPM packings with $N=64$ as a function of
asphericity ${\cal A}$. In (a) and (c), ${\cal A}$ is normalized by 
the area ${\cal A}_v$ of a regular polygon with $N_v$ vertices. 
For monodisperse systems with the smooth surface model, $N_{v} = 12$
(squares), while $N_{v} = 12$ (circles), $24$ (triangles), and $34$
(stars) for monodisperse systems with the rough surface model. Bidisperse 
systems (exes) have $N_v = 17$ ($12$) for the large (small) polygons, 
using the rough surface model. The dashed lines in (a) and (b) 
indicate 
${\cal A} = {\cal A}^* \approx 1.16$ at which packings become confluent 
in the large $N_v$ limit. 
In (a), we also show $\phi_J/\phi_{\rm max} \approx
0.81$ (with $\phi_{\rm max}=1$) for $N=64$ monodisperse, frictional discs using the
Cundall-Strack model with $\mu=0.65$ (filled diamond).  In 
(c), the dashed line indicates 
$z({\cal A}/{\cal A}_v)=z(1)+z_0({\cal A}/{\cal A}_v-1)^{\beta}$, where 
$z(1) \approx 3.3$, $\mu=0.65$, $z_0 \approx 3.9$, and $\beta \approx 0.25$.}
\label{fig.2}
\end{figure}
 
In Fig.~\ref{fig.2} (c), we show the coordination number $z$ versus
${\cal A}/{\cal A}_v$ for $N=64$ deformable polygons for both surface
roughness models.  Near ${\cal A}/{\cal A}_v =1$, the smooth
model yields packings with $z \approx 4$ (where rattler polygons with
fewer than $2$ interparticle contacts are not included). This result
is consistent with isostatic packings~\cite{tkachenko} of
frictionless, monodisperse and bidisperse disks. In contrast, $z < 4$
near ${\cal A}/{\cal A}_v =1$ using the rough surface model, which is
consistent with studies of packings of frictional
disks~\cite{silbertfric,ohernfriction}. For both roughness
models, $z({\cal A}/{\cal A}_v)-z(1)$ increases as a power-law in
${\cal A}/{\cal A}_v-1$. We find that $z=5.8 \pm 0.1$ at confluence
when ${\cal A} = {\cal A}^*$.  In contrast, prior work has suggested
that $z=5$ is the isostatic contact number for the vertex
model~\cite{binature}.

We also measured the effective friction coefficient $\mu_c =|{\bf
  F}_{mn}^t|/ |{\bf F}_{mn}^r|$ at each contact $c$ between polygons
$m$ and $n$ in DPM packings using the rough surface model. $|{\bf
  F}_{mn}^r|$ ($|{\bf F}_{mn}^t|$) is the normal (tangential)
component of the repulsive contact force. For each packing, we find
the maximum $\mu_c$ over all contacts and average it over at least
$500$ packings. The average maximum friction coefficient $\mu$ depends
on $N_v$ and $l_0$ in the rigid polygon limit (${\cal A}={\cal
  A}_v$). For $N_v=12$ and $l_0=1$, $\mu \approx 0.7$ for ${\cal
  A}_{12} \approx 1.02$ and decreases as $N_v$ increases. In
Fig.~\ref{fig.2} (d), we show that $\mu$ increases by an order of
magnitude as ${\cal A}$ increases from $\approx 1$ to $1.25$. We find
similar increases for $\mu({\cal A})$ when using different 
$N_v$.  Despite the strong increase in $\mu$ for the rough surface
model, both the smooth and rough models yield similar results
for $\phi_J({\cal A})$ and $z({\cal A})$ away from the rigid-disk
limit. Thus, particle deformation weakens the influence of friction on
the structural properties of jammed DPM packings.

\begin{figure*}[ht]
\captionsetup[subfloat]{labelformat=empty,justification=raggedright ,singlelinecheck=false}
\includegraphics[trim={0.5cm 0.5cm 0.5cm 1cm},clip, width=0.90\textwidth]{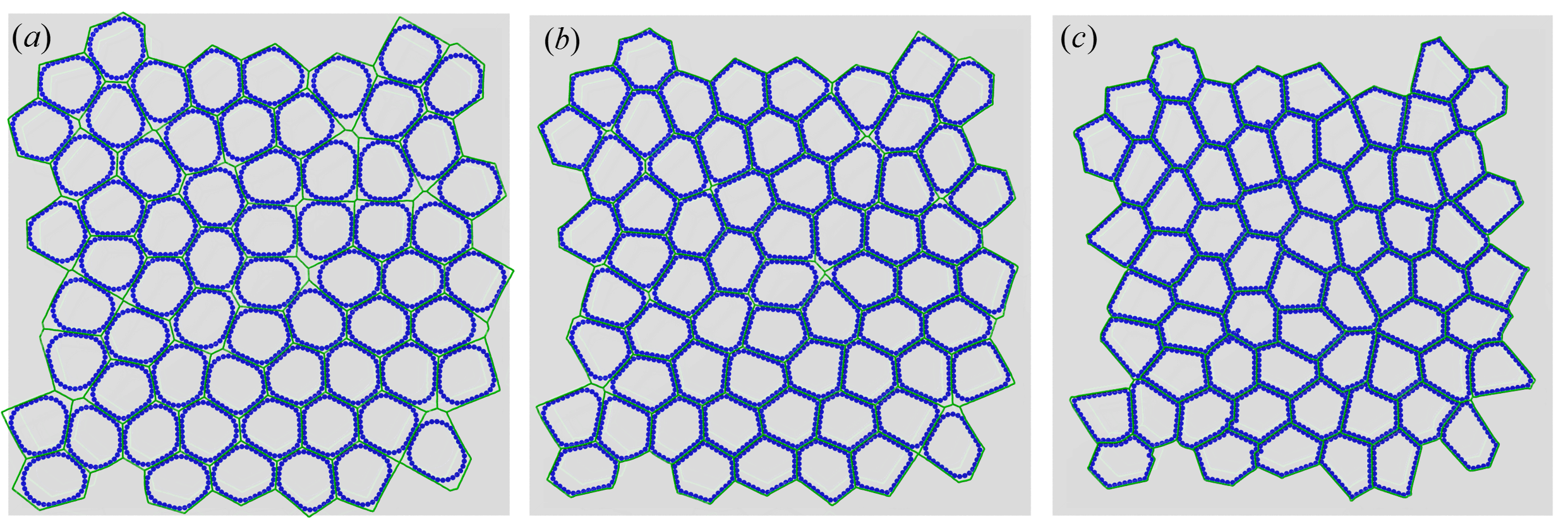}\label{fig.3b}
\caption{Jammed DPM packings for the rough surface 
model with $N_v=34$ and (a) ${\cal A}=1.03$, (b) $1.08$, and (c)
$1.16$, near ${\cal A}^*$. The polygonal cells (solid lines) 
surrounding each DPM are 
obtained from a surface-Voronoi tessellation.}
\label{fig.3}
\end{figure*}

To understand the value ${\cal A}^* \approx 1.16$ above which DPM
packings are confluent, we calculate the free
area versus ${\cal A}$ using surface-Voronoi
tessellation~\cite{schaller}.  In Fig.~\ref{fig.3}, we show example
packings at three values of ${\cal A}$ approaching ${\cal
  A}^*$. At ${\cal A}=1.03$, well-below ${\cal A}^*$, the deformable
polygons are quasi-circular and there is a relatively large amount of
free area. As ${\cal A}$ increases, the ``effective'' sides of the deformable
polygons straighten and fill the surface-Voronoi cells. When ${\cal
  A} \sim {\cal A}^*$, it is difficult to differentiate the DPM
polygons from the surface-Voronoi cells. 

\begin{figure}[h]
\includegraphics[trim={0.25cm 0cm 0.25cm 0.1cm},clip, width=0.92\textwidth]{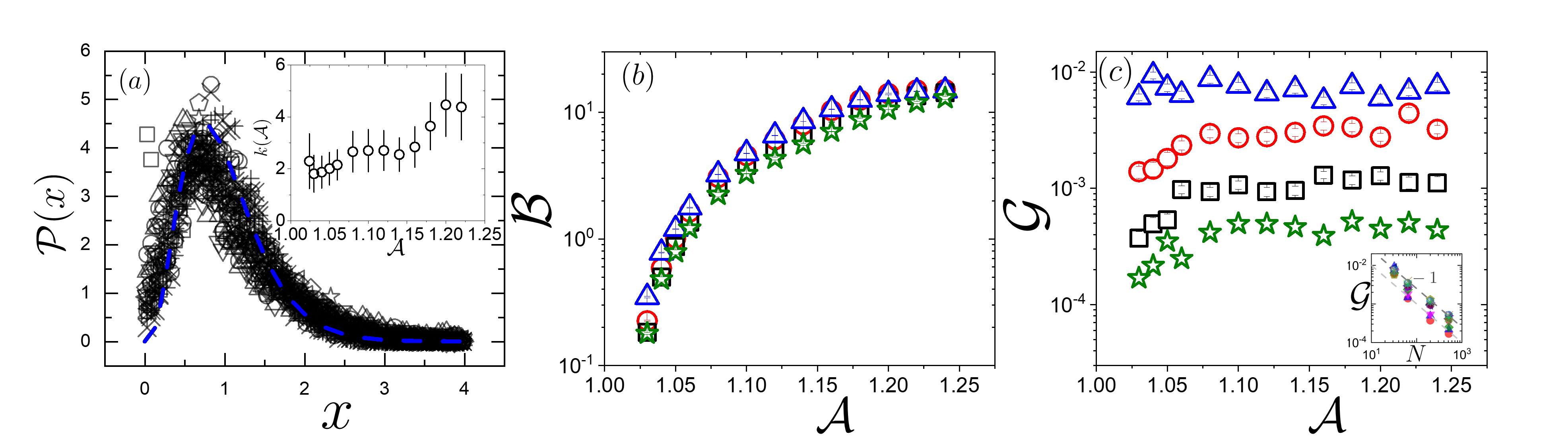}
\caption[]{(a) Distribution of areas $a_t$ of the surface-Voronoi tessellated polygons for $14$ values of the asphericity from ${\cal A} = 1.02$ (squares) to 
$1.25$ (exes) for $N=64$ monodisperse DPM polygons with $N_v=12$ and the rough 
surface model. The distributions ${\cal P}(x)$ are plotted against the 
rescaled variable $x=(a_t-a_{min})/(\langle
a_t \rangle - a_{min})$, where $a_{min}$ is the minimum tessellated area 
for each packing.  (inset) ${\cal P}(x)$ resemble $k$-gamma distributions 
with $k$-values that depend on ${\cal A}$. (b) Bulk ${\cal
B}$ and (c) shear ${\cal G}$ moduli for jammed DPM packings using the model 
in (a) versus ${\cal A}$ for 
system sizes $N= 32$ (triangles), $64$ (circles), $200$ (squares), and
$512$ (stars). The inset to (c) shows the system-size scaling of 
${\cal G}$. The dashed lines have slope $-1$.}
\label{fig.4}
\end{figure}

Prior studies have shown that the areas of Voronoi polygons for
hard-disk configurations follow $k$-gamma
distributions~\cite{kumar_jcp,aste_pre_emergence}, which can be written as
\begin{equation}
\label{kgamma}
{\cal P}(x) =\frac{k^{k}}{(k-1)!}x^{k-1}\exp(-kx),
\end{equation}
where $x=(a_{t}-a_{min})/(\langle a_{t} \rangle - a_{min})$, $a_t$ is
the area of each Voronoi polygon, $a_{min}$ is the area of the
smallest Voronoi polygon, $\langle a_t\rangle$ is an average over
Voronoi polygons in a given system, $k=(\langle a_t
\rangle-a_{min})^2/\sigma_a^2$, and $\sigma_a^2 = \langle (a_t -
a_{min})^2\rangle$ controls the width of the distribution.  In
Fig.~\ref{fig.4} (a), we show that the distribution ${\cal P}(x)$ for
DPM packings resembles a $k$-gamma distribution with a $k$-value that
depends on ${\cal A}$. The inset shows that $k$ increases from $2$ to
$\approx 5$ over the range $1 < {\cal A} < 1.25$. Prior studies have
shown similar values for $k$ for Voronoi-tessellated hard
disks~\cite{kumar_jcp} ($k=3.6$) and jammed bidisperse
foams~\cite{vanhecke_epl} ($k \approx 6$).

In Fig.~\ref{fig.4} (b) and (c), we show the bulk ${\cal B}$ and shear
${\cal G}$ moduli for DPM packings (rough surface model with $N_v=12$)
versus ${\cal A}$ for several $N$. ${\cal B}$ is roughly independent
of $N$ and grows strongly with ${\cal A}$ (changing by more than two
orders of magnitude) as packings gain contacts. In contrast, at each
$N$, the shear modulus ${\cal G}$ increases only by a factor of $3$ as
${\cal A}$ increases from $1$ to $1.25$. As a result, the ratio ${\cal
  B}/{\cal G}$ varies from $10^{3}$ to $10^{5}$, indicating that the
system is in the isotropic elastic limit, over this range of ${\cal
  A}$~\cite{poissonnature}. The inset of Fig.~\ref{fig.4} (c) shows
that even though DPM packings are solid-like with non-zero shear
moduli ${\cal G} > 0$ for any finite $N$, ${\cal G}$ scales as
$N^{-1}$ with increasing system size. Similar scaling was found for
${\cal G}$ in jammed disk packings~\cite{goodrich_2012_finite}. Disk
as well as DPM packings can be stabilized in the large-system limit by
adding nonzero pressure.

\begin{figure}
\includegraphics[trim={0cm 0.0cm 0cm 0cm},clip, width=1.0\textwidth]{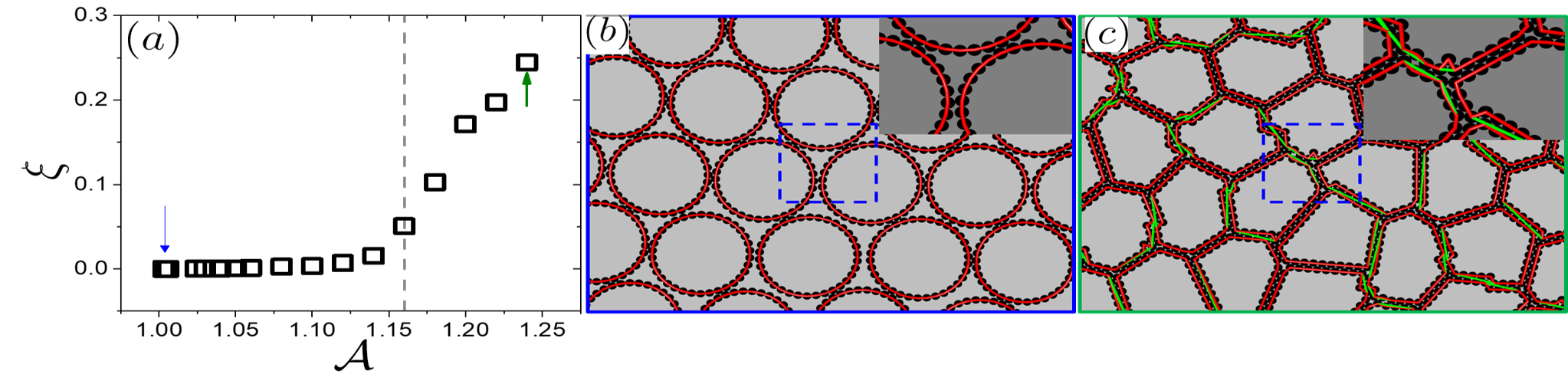}
\caption{(a) Excess perimeter $\xi =
p - p_{conv}$ of DPM packings, where $p$ is the DPM perimeter and $p_{conv}$ is the perimeter of the DPM 
convex hull (rough surface 
model with $N_v =34$) plotted versus ${\cal A}$.  The vertical dashed 
line indicates ${\cal A}^* \approx 1.16$ and blue and green arrows
indicate the values of ${\cal A}$ for packings in (b) and (c), 
respectively. The red and yellow solid lines represent perimeters of the 
DPM and convex hull, respectively. The insets in (b) and (c) are close-ups 
of the regions indicated by blue dashed boxes.}
\label{fig.5}
\end{figure}

 The coordination number and bulk and shear moduli vary
continuously as ${\cal A}$ increases above ${\cal A}^*$. Other than
being confluent for ${\cal A} > {\cal A}^*$, what is different about
DPM packings for ${\cal A}$ above versus
below ${\cal A}^*$?  In Fig.~\ref{fig.5} (a), we show the excess
perimeter $\xi = p - p_{conv}$ for DPM packings,
where $p_{conv}$ is the perimeter of the convex hull of each
$N_v$-sided polygon~\cite{liu_2015_optimizing}. $p \approx
p_{conv}$ (with $\xi = 0$) for ${\cal A} < {\cal A}^*$ as shown in
Fig.~\ref{fig.5} (b). $\xi$ becomes nonzero for ${\cal
  A} > {\cal A}^*$ when the deformable polygons buckle and develop
invaginations [Fig.~\ref{fig.5} (c)].  Thus,
DPM packings at confluence are under tension for
${\cal A} < {\cal A}^*$ and under compression for ${\cal A} >
{\cal A}^*$.\\
We developed the DPM model, which can be used to study 2D cellular
materials composed of deformable particles, including foams,
emulsions, and cell monolayers, over a range of packing fraction,
particle shape and deformability. We showed that the packing fraction
at jamming onset $\phi_J$ grows with particle asphericity, ${\cal A}$,
reaching confluence at ${\cal A}^* \approx 1.16$. ${\cal A}^*$
coincides with the value of the asphericity at which DPM polygons
fill the cells from the surface-Voronoi tessellation of DPM
packings.  By calculating their shear modulus ${\cal G}$, we show that
DPM packings are solid-like above and below ${\cal A}^*$.  For ${\cal
  A} > {\cal A}^*$, DPM polygons possess invaginations that grow with
${\cal A}-{\cal A}^*$. Thus, at confluence, DPM packings are under
compression for ${\cal A} > {\cal A}^*$ and under tension for ${\cal
  A} < {\cal A}^*$. In future studies, we will extend the DPM to 3D to investigate the material properties of
tissues. Based on Voronoi tessellations of sphere
packings~\cite{klatt}, we expect that DPM packings in 3D will be
confluent for ${\cal A}_{3D} > {\cal A}_{3D}^* \approx 1.18$~\cite{manning}, where
${\cal A}_{3D}=s^{3/2}/6\sqrt{\pi}v$, and $s$ and $v$ are the surface
area and volume of the DPM particles.
\onecolumngrid

\twocolumngrid

\begin{acknowledgments}
We acknowledge support from NSF Grant Nos. PHY-1522467
(A.B.), CMMI-1462439 (C.O.) and CMMI-1463455 (M.S.),
the President's International Fellowship Initiative (PIFI) and Hundred-talent Program of Chinese Academy of Sciences (A.B. and F.Y.)
, and National Library of
Medicine Training Grant T15LM00705628 (A.S.). This work was also
supported by the High Performance Computing facilities operated by,
and the staff of, the Yale Center for Research Computing.
\end{acknowledgments}

\balance
\nocite{*}
\bibliographystyle{apsrev4-1}
%

\end{document}